\documentclass[aps,prc,twocolumn,floatfix]{revtex4}

\usepackage{epsfig}

\begin{document}

\title{Azimuthal flow of decay photons in  
	 relativistic nuclear collisions}
\author{Biswanath Layek}
\affiliation{Variable Energy Cyclotron 
Centre, 1/AF Bidhan Nagar, Kolkata 700 064, India}            
\author{Rupa Chatterjee}
\affiliation{Variable Energy Cyclotron 
Centre, 1/AF Bidhan Nagar, Kolkata 700 064, India}            
\author{Dinesh K.~Srivastava}
\affiliation{Variable Energy Cyclotron 
Centre, 1/AF Bidhan Nagar, Kolkata 700 064, India}            
\date{\today}

\begin{abstract}
An overwhelming fraction of photons from relativistic heavy ion collisions
has its origin in the decay of $\pi^0$ and $\eta$ mesons.
 We calculate the azimuthal
asymmetry of the decay photons for several azimuthally asymmetric pion 
distributions.  We find that the $k_T$ dependence of the elliptic flow parameter
$v_2$ for the decay photons closely follows the
elliptic flow parameter $v_2^{\pi^0}$ evaluated at
 $p_T \approx k_T+\delta$, where 
$\delta\approx$ 0.1 -- 0.2 GeV, for typical pion distributions measured in
nucleus-nucleus collisions at relativistic energies. Similar results are
obtained for photons from the 2-$\gamma$ decay of $\eta$ mesons.
Assuming that the flow of $\pi^0$ is similar to those for $\pi^+$ and $\pi^-$
for which independent measurements would be generally available, this
ansatz can help in identifying additional sources for 
photons. Taken along with quark number scaling 
suggested by the recombination model, it may help to estimate
$v_2$ of the parton distributions in terms of 
azimuthal asymmetry of the decay photons at large $k_T$.
\end{abstract}

\maketitle
\section{Introduction}

The azimuthal flow of particles produced in relativistic heavy ion
collisions has provided  a strong evidence for the creation of a hot
and dense system very early in non-central collisions~\cite{v2_exp}. 
The importance of this observation stems from the fact that pressure
gradients generated in the system at very early times, transform the 
eccentricity in co-ordinate space for such collisions to the momentum
space for distribution of the produced particles~\cite{v2_theo}.
The $p_T$ dependence of the elliptic flow parameter $v_2$ at lower transverse
momenta has been quantitatively explained using hydrodynamics~\cite{peter},
The observed saturation of $v_2$ at higher $p_T$ has been attributed to 
effects of viscosity~\cite{derek} or incomplete thermalization~\cite{incomplete}.
 The  observed scaling of $v_2$ with the number of valence quarks
in the hadrons is understood in terms of the recombination
model for hadronization~\cite{reco1,reco2}. 

It has recently been suggested that the study of the elliptic flow of thermal 
photons~\cite{phot_flow} may provide valuable insight into the build-up of
azimuthal flow with time. 
This will require a subtraction of the contribution of photons from
the decay of $\pi^0$ and $\eta$ mesons produced in the 
collisions~\cite{phenix2}.

The deviation of elliptic flow of inclusive photons from that of
 decay photons, will thus confirm the presence of additional sources 
of photons. We shall see later that the
elliptic flow of photons from the decay of $\pi^0$ and $\eta$ mesons
  may also provide useful estimates for $v_2$ of the partons in the 
frame-work of the recombination model.

In the present work, we suggest an ansatz for the evaluation of the 
elliptic flow of photons from the decay of pions, which could be useful
for such studies. In the next section we study the transverse momentum
dependence of the elliptic flow parameter for decay photons for several
momentum distribution functions for the $\pi^0$ and $\eta$ mesons.
In section III; we evaluate the $v_2$ for photons arising from 2-$\gamma$ 
decay of $\pi^0$ and $\eta$ mesons, which are formed from 
recombination of partons to get a
direct relation with $v_2$ for partons. Finally, we
give our conclusions.

\section{Photons from decay of $\pi^0$ and $\eta$ mesons}

Consider a $\pi^0$ having  four-momentum $p$ and mass $m$ decaying into two
photons. The momentum distribution of photons 
in an invariant form is given by~\cite{cahn}:
\begin{equation}
k_0\frac{dN}{d^3k}(p, k)=\frac{1}{\pi}\, \delta(p \cdot k-\frac{1}{2}
m^2)\, ,
\label{dec}
\end{equation}
where $k$ is the four-momentum of the photon.

Thus, the Lorentz invariant cross-section for photon production is

\begin{equation}
k_0\frac{d\sigma}{d^3k}=\int \frac{d^3p}{E}\left(E \frac{d\sigma}{d^3p} \right)
\frac{1}{\pi} \delta (p \cdot k -\frac{1}{2} m^2) \, .
\label{dsig}
\end{equation}

Let us write the pion distribution as:

\begin{equation}
\frac{d\sigma}{d^2p_T \, dy}=\frac{d\sigma}{2 \pi p_T \, dp_T \, dy}
\left[1+2 v_2(p_T) \cos(2 \phi)+...\right]
\label{azi}
\end{equation}
for  a non-central collision of identical nuclei, where $v_2(p_T)$ is the
momentum dependent elliptic flow parameter. We consider two typical 
parameterizations for the pion distribution: an exponential distribution
and a power-law distribution. For the first case, we assume
\begin{equation}
 \frac{d\sigma}{p_T \, dp_T \,dy} \sim \exp( -\sqrt{p_T^2+m^2}/T_0 )
\times \exp \left( -y^2/2 \alpha \right ) 
\label{exp}
\end{equation}
where the slope parameter $T_0$ is of the order of 290 MeV~\cite{star}
 and the 
width parameter of the rapidity distribution $\alpha$ is taken as $\sim$~4.
For the power-law distribution, we take~\cite{wa98}
\begin{equation}
 \frac{d\sigma}{p_T \, dp_T \,dy} \sim \left ( \frac{p_0}{p_0+p_T} \right)^n
\times \exp \left( -y^2/2 \alpha \right ) 
\label{power}
\end{equation}
where $p_0 \sim$ 5 GeV and $n$ is about 29. The elliptic flow parameter $v_2$
is approximated as
\begin{equation}
v_2(p_T)=a\left[ 1-\exp(-p_t/b)\right]
\label{v2}
\end{equation}
which approximately reflects the increase of $v_2$ with $p_T$ for
lower transverse momenta and saturation of its value at larger $p_T$. We have
taken $a=$ 0.2 and $b =$ 1~GeV for the first set of calculations.
All the results given in the present work are for photon rapidity equal to
zero.
\begin{figure}
\centerline{\epsfig{file=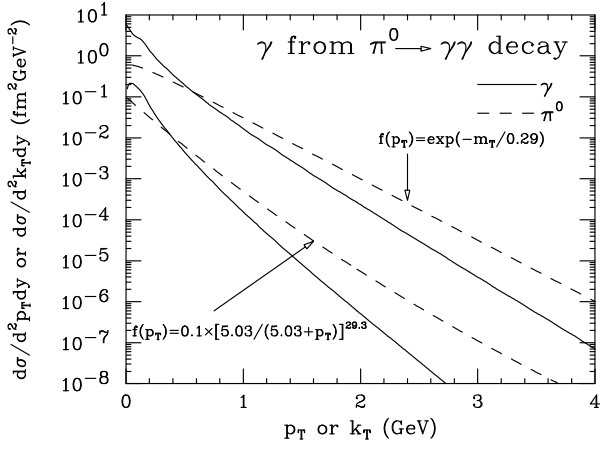,width=7.9cm}}
\vskip 0.2in
\centerline{\epsfig{file=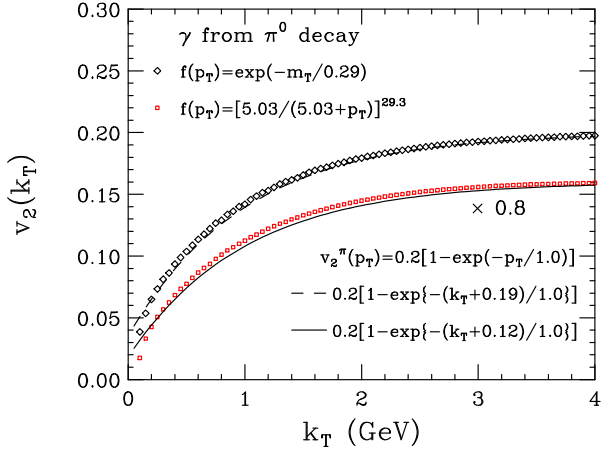,width=7.9cm}}
\caption{(Color on-line) Upper Panel: Spectrum of photons from the decay of $\pi^0$
for an exponential (Eq.~\ref{exp}) and a power-law (Eq.\ref{power}) 
distribution for pions.
Lower Panel: Elliptic flow of photons from decay of $\pi^0$ having 
elliptic flow given by Eq.~\ref{v2} at $y=0$. $f(p_T)$ stands
for the momentum distribution of the $\pi^0$.  The symbols 
give the result of numerical calculation, while the curves give the 
fits.
}
\label{fig1}
\end{figure}

Eq.~\ref{dsig} is then evaluated numerically and $v_2(k_T)$ for the resulting photon
distribution obtained. In Fig.~\ref{fig1} we give our results
for the spectrum of photons from decay of $\pi^0$ along with the variation
of transverse momentum dependence of their elliptic flow parameter. We find that for both
distributions, the resulting $v_2(k_T)$ for the decay photons can be
approximated as:

\begin{equation}
v_2(k_T)\approx v_2^{\pi^0}(p_T)
\label{pigam}
\end{equation}
where
\begin{equation}
p_T \approx k_T+\delta
\label{del}
\end{equation}
and $\delta \approx$ 0.1 -- 0.2 GeV, for $k_T >$ 0.2 GeV, to an accuracy of better than
1 --3 \%. 

\begin{figure}
\centerline{\epsfig{file=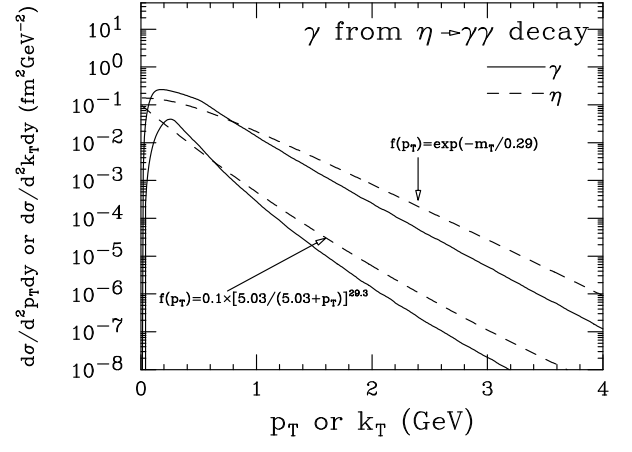,width=7.9cm}}
\vskip 0.2in
\centerline{\epsfig{file=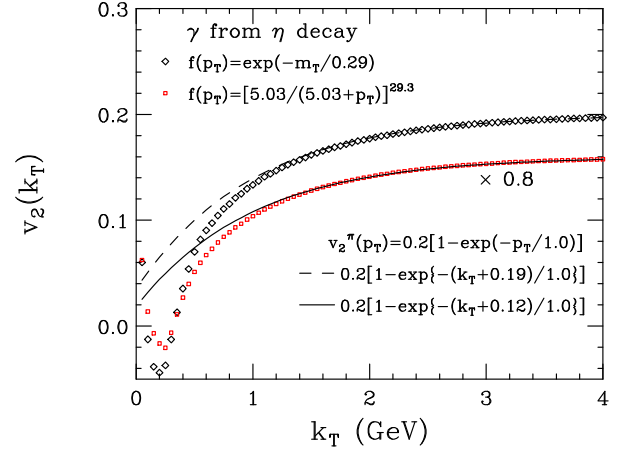,width=7.9cm}}
\caption{(Color on-line) Same as Fig.~\ref{fig1} for $\eta$ mesons.}
\label{fig2}
\end{figure}

This result can be understood as follows.  The $\delta$-function in
 Eq.~\ref{dec} provides that
\begin{equation}
m_T k_T \cosh (y_p-y_k)-p_T k_T \cos(\phi_p-\phi_k) -\frac{1}{2} m^2 = 0~,
\end{equation}
in an obvious notation. Next, we note that once the momentum of the pions is 
large, the opening angle for the decay-photons will be very small~\cite{cahn}.
Thus,  for example, in the extreme case,  we have,
$y_p-y_k \approx 0$ and $\phi_p-\phi_k \approx 0 ~{\rm or}~\pi$,  for the
photons  which are co-linear with the pion. This leads to the solutions:
 $k_T \approx p_T$ and $k_T \approx (m^2/4p_T)(1-m^2/4p_T^2)$ for large $p_T$.

However, in general, the two photons between them will 
cover the entire range of the  allowed transverse momentum.
Both the photons, which will be 
almost co-linear with the pion, will ``inherit" the $v_2$ of the pion,
which has a larger transverse momentum.
The photon with the lower $k_T$ will, however, be submerged in a much larger
yield of photons coming from  decay of $\pi^0$ having lower $p_T$, as 
the transverse momentum distribution of pions falls steeply with
increase in $p_T$.
Thus, in general the $v_2$ of a photon with a given transverse 
momentum will be larger than the $v_2$ of a pion with the same
transverse momentum. This accounts for the shift, Eq.~\ref{del}.

This argument will hold even if the $v_2(p_T)$ for 
the pions decreases with increase in $p_T$, which is likely for
larger $p_T$. 
The only difference would be that $v_2$ for the
decay photons will now be smaller than the $v_2$ for the
pions at the same transverse momentum (see later; Fig.~\ref{fig5}). 

The spectra and the transverse momentum dependence of the 
elliptic flow parameter for the $\eta$ mesons are given in Fig.~\ref{fig2}. 
The rich structure seen for the $\eta$ mesons at smaller transverse momenta has its 
origin in the large mass of the $\eta$ mesons. Thus, we find that the
 $v_2$ for decay photons coming from the 2-$\gamma$ decay of $\eta$ mesons, 
approaches the $v_2$ for the mesons only at larger
transverse momenta ($>$ 0.8 -- 1.0 GeV). For the
simple parameterizations of the momenta considered here,
$v_2$ is negative for very low $k_T$ as the decay photons are
distributed away from the major axis in the momentum space
when the $p_T$ of the $\eta$- mesons is low and the opening angles are large.

Of-course, a more complete analysis would include
the photons coming from the 3-$\pi^0$ decay of the $\eta$ mesons, which 
will mainly populate the low $k_T$ window. While a simulation of the
contribution of this process to the $v_2$ can be straight-forward
using standard event generators, a direct evaluation will not be easy
as it would involve a 10-dimensional numerical integration. 

However, we can easily make an estimate of the range of the transverse 
momenta of photons, where the photons arising from decay 
$\eta \rightarrow 3 \pi^0 \rightarrow 6 \gamma$ would contribute. 
The maximum kinetic energy of a $\pi^0$ in the rest frame of the
$\eta$ meson is given by ~\cite{khuri}:
\begin{equation}
T=\frac{(m_\eta-m_\pi )^2 -4 m_\pi^2}{2 m_\eta}\approx \, {\rm 80 \, MeV}.
\label{ke}
\end{equation} 

Thus, for example, the maximum energy of a pion emerging from an $\eta$ meson 
having a momentum of 2 GeV would be of the order of 0.62 GeV and would
contribute to photons having momenta less than that. Next, we recall that
$\eta/\pi^0$ is of the order of 0.44 for $p_T >$ 2 GeV and 
is expected to drop to zero at $p_T =0$, at least according to calculations
based on PYTHIA~\cite{etapi}. This coupled with the branching 
ratio of about 0.3 for this mode of decay should ensure that the contribution
of photons from the 3-$\pi^0$ decay of $\eta$ mesons would be limited to 
smaller momenta. A more detailed study of this effect would definitely be useful. 

\begin{figure}
\centerline{\epsfig{file=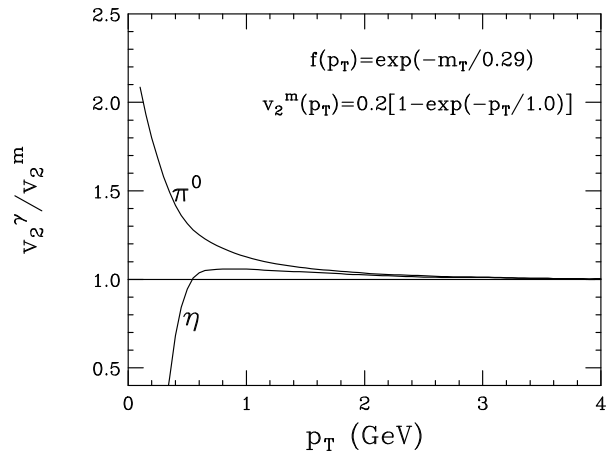,width=7.9cm}}
\caption{Transverse momentum dependence of the ratio of elliptic flow
parameters for decay photons and the $\pi^0$ or $\eta$ meson.
}
\label{fig3}
\end{figure}

Let us return to the discussion of the $v_2$ of photons from the
 $2 \gamma$ - decay of $\pi^0$ and $\eta$ mesons. 
We now plot the ratio of the elliptic
flow parameters of the calculated decay photons and the mesons as a 
function of their transverse momenta in Fig.~\ref{fig3}. We see once again
that $v_2$ for the two are quite similar when transverse momenta exceeds 1 GeV.
Similar results for the
transverse momentum dependence of
 $v_2^{\gamma}/v_2^{\pi^0}$ had earlier been noted by the
 WA98 group~\cite{wa98_flow}.
 We now understand this in terms of the momentum shift
 $\delta$ (Eq.~\ref{del}).
 
\begin{figure}
\centerline{\epsfig{file=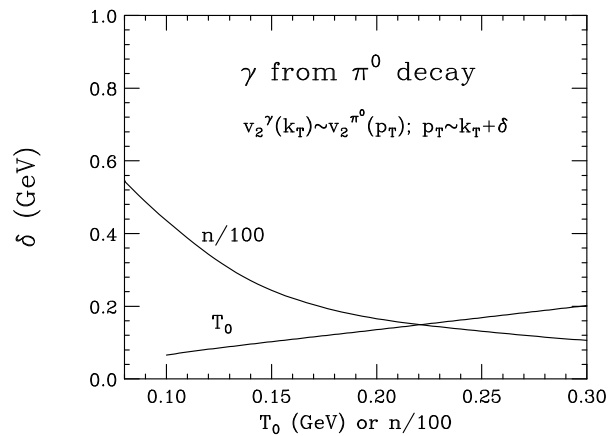,width=7.9cm}}
\caption{ Variation of the shift parameter $\delta$ with the slope 
parameter ($T_0$) and the power-law parameter $n$ for the photons
coming from the decay of $\pi^0$.}
\label{fig4}
\end{figure}

\begin{figure}
\centerline{\epsfig{file=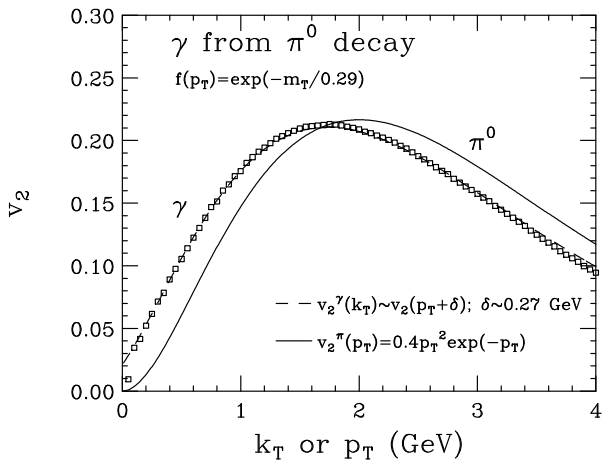,width=7.9cm}}
\caption{Elliptic flow parameter for photons from the 2-$\gamma$ 
decay of $\pi^0$. The symbols denote the calculated $v_2$ while
the dashed curve denotes the fit.  The $v_2 (p_T)$  for pions
 is taken to rise first, reach a maximum and drop for
larger transverse momenta.}
\label{fig5}
\end{figure}

Considering the likely usefulness of the relation given by
Eq.~\ref{del} we have examined it to determine the dependence of
$\delta$ on the slope-parameter $T_0$ or the power-law parameter $n$ (see
Fig.\ref{fig4}). We see that, in general,  the shift $\delta$ is smaller
 if the distribution is steeper. This may be
considered as a generalization of the Sternheimer's~\cite{stern}
prescription for azimuthally asymmetric distribution of 
pions and their decay photons.

 We now take a more general 
shape of the transverse momentum dependence of the $v_2$ for pions,
which rises with $p_T$, reaches a maximum, and then starts decreasing.
We illustrate our discussion (see Fig.\ref{fig5}) by taking,
\begin{equation}
v_2(p_T)=0.4 \, p_T^2 \, \exp(-p_T) \, .
\label{peak}
\end{equation}
We again see that the transverse momentum dependence of
 $v_2$ for the decay photons is obtained
from the $v_2$ for the pions by a shift of about 0.27 GeV.


\begin{figure}
\centerline{\epsfig{file=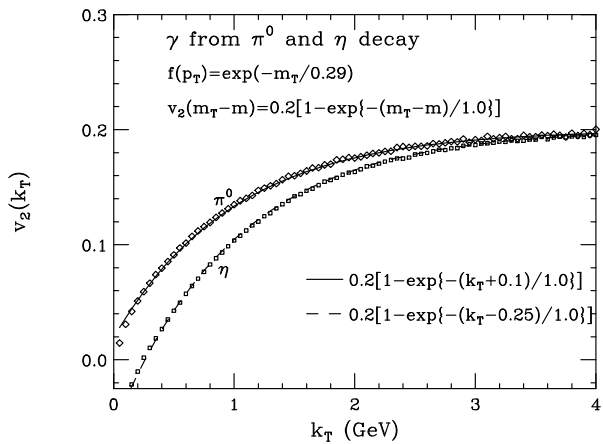,width=7.9cm}}
\caption{Elliptic flow parameter for photons from the 2-$\gamma$ 
decay of $\pi^0$
and $\eta$ mesons. The hadron $v_2$ is taken to scale with "transverse
kinetic energy" (Ref.~\cite{mt}). Symbols, as before, denote the
calculated values.}
\label{fig6}
\end{figure}

 It has been recently reported~\cite{mt} that the so-called
quark-number scaling of the elliptic flow parameter suggested by the
recombination model works much more accurately when plotted in terms of
``transverse kinetic energy" or $m_T-m$, where $m$ is the mass of the
hadron. In order to see the consequences of this behaviour of the elliptic
flow parameter, we now replace $p_T$ with $m_T-m$ in Eq.~\ref{v2}. Next,
taking the exponential distribution function (Eq.~\ref{exp}) for
transverse momenta, we calculate the elliptic flow parameter for
decay photons from $\pi^0$ and $\eta$ mesons (see Fig.~\ref{fig6}).

Identifying $k_T$ of the photons with the so-called ``transverse kinetic
energy", we again see that $v_2$ for photons is simply related to
$v_2$ for the pions by a small shift of $\approx$ 0.1 GeV in
the transverse kinetic energy of $\pi^0$. 
 We also find that we can easily describe the $v_2$ of the photons from the 
decay of $\eta$ mesons by a small negative shift.

\begin{figure}
\centerline{\epsfig{file=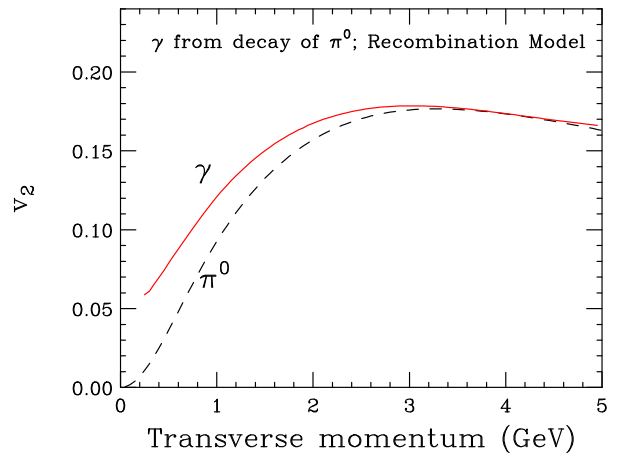,width=7.9cm}}
\caption{(Color on-line) Elliptic flow parameter for photons (solid curve)
 from the decay of $\pi^0$ (dashed curve)
 obtained using the recombination model.}
\label{fig7}
\end{figure}

Thus, we see that the elliptic flow parameter of
decay photons from $\pi^0$ which make an overwhelming
contribution to the inclusive photons in relativistic heavy ion
collisions is simply related (Eq.\ref{del})
to the elliptic flow parameter for $\pi^0$.
As we can get independent information about the later in 
terms of the flow of $\pi^+$ and $\pi^-$, these results could
be useful in identifying additional sources of photons~\cite{phenix2}.

\section{Recombination model and decay photons} 

As remarked earlier, one of the more interesting results of 
elliptic flow studies has been
the approximate quark-number scaling of the hadronic $v_2$.
This finds a natural
explanation in terms of the recombination model for hadronization 
of the quark gluon plasma~\cite{reco1,reco2}, which suggests that
in the region of $p_T$ where the recombination of the partons 
dominates the process of hadronization, that is $p_T <$ 4 (6) GeV
for mesons (baryons) and the effect of mass is minimal, $v_2$ obeys
a simple scaling law:
\begin{equation}
v_2(p_T)\approx n v_2^{q}(p_T/n)
\label{scale}
\end{equation}
where $n$ is the number of valence quarks in the hadron and $v_2^q$ is the
elliptic flow parameter for them.

\begin{figure}
\centerline{\epsfig{file=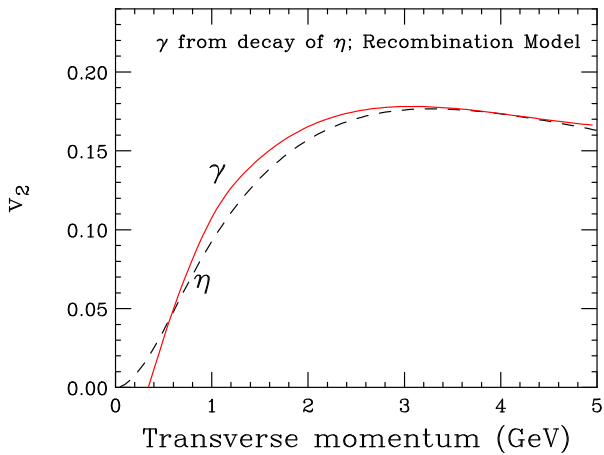,width=7.9cm}}
\caption{ (Color on-line) Elliptic flow parameter for photons (solid curve)
 from the 2-$\gamma$  decay of $\eta$-mesons (dashed curve) obtained 
using the recombination model.}
\label{fig8}
\end{figure}

The shifted-scaling of the $v_2$ 
of the decay photons with the $v_2$ for the $\pi^0$ and $\eta$ seen
earlier for a variety of momentum distributions and momentum-dependences of
$v_2$, taken along with the valence-quark scaling of the
$v_2$ for hadrons, holds out the promise of the ``lowly" decay photons providing a
direct measure of the $v_2^q$ of the partons. In order to examine
this possibility, we now explicitly calculate the $v_2$ for photons from
the decay of $\pi^0$ and $\eta$, which in turn, are formed by the
recombination of partons. 

The transverse momentum distribution of mesons formed by the
recombination from thermal parton distribution can be written as~\cite{reco2}:
\begin{eqnarray}
\frac{dN}{d^2p_T dy}(y=0) &\sim&
K_1\left[\frac{2 (m_q^2+p_T^2/4)^{1/2} \cosh \eta_T}{T}\right]
 \nonumber\\
&& \times
 m_T I_0\left[\frac{p_T\sinh \eta_T}{T}\right]~.
\label{rainer}
\end{eqnarray} 
In the above $m_T$ is the transverse mass of the hadron, $m_q = $260 MeV is the 
quark mass, $T=$ 175 MeV is the  transition 
temperature, and $\eta_T$ is the transverse
rapidity such that $\tanh \eta_T =0.55 $,
appropriate for a system created in collision of gold
nuclei at $\sqrt{s_{NN}}$ =200 GeV. While writing the
above we have utilized the $\delta$-function for the asymptotic form of
the perturbative meson distribution amplitude.

 We now return to the present calculation. In these illustrative studies,
 we account for the elliptic flow of the mesons  by multiplying the
above distribution with $[1+4v_2^{q}(p_T/2)\cos(2\phi)]$. 
We have calculated $v_2^{q}(p_T)$, the
elliptic flow for the partons, using Eq.~81 of Fries {\em et al.}~\cite{reco2}.

In Fig.~\ref{fig7}, we give our results for the elliptic flow
parameter for pions obtained, using
the recombination model along with the $v_2$ for the photons coming
from the decay.  We should not take the results for the $v_2$ for pions
for lower $p_T$
too literally, as the pion mass is much smaller than $2m_q$ which goes
into making it.  However, for larger $p_T$ this difference is
not very relevant and we can trust the
model. We see that $v_2^{\gamma}$ closely follows the $v_2^\pi$ as before.

The mass of $\eta$ mesons on the other hands is large enough to be  free from
this complication. We see that once again, the decay photons closely
follow the azimuthal asymmetry of the $\eta$ mesons at larger $p_T$ (see
Fig.~\ref{fig8}).

\section{Summary}

We have studied the azimuthal asymmetry of distribution of photons 
originating from the 2-$\gamma$ decay of $\pi^0$ and $\eta$ mesons produced in
relativistic heavy ion collisions. Several 
distribution functions and parameterizations of the transverse momentum
dependence of the elliptic flow parameter, including those coming
from the recombination model have been considered. 

We have empirically found that the elliptic flow parameter for the
photons from the decay of pions is quite close to the same for the
pions evaluated at a transverse
momentum shifted by about 0.1 -- 0.2 GeV. This has its
origin in the small opening angle for the photons for pions having
large momenta.
Similar results are found for photons from the 2-$\gamma$ 
decay of $\eta$ mesons, at $p_T >$ 1 GeV.

These empirical findings could be useful in identifying additional sources
of photons, as the elliptical flow of $\pi^0$ should be similar to that for
$\pi^+$ and $\pi^-$. The flow of $\eta$ mesons could perhaps be approximated
with those for kaons based on considerations of their masses and the number of
valence quarks. In any case, in general,
the yield of eta mesons is much less compared to that for pions. Thus the
presence of an additional source of photons will be indicated by
 the deviation of the
$v_2$ for inclusive photons from the $v_2$ of decay photons from the
2-$\gamma$ decay of $\pi^0$ and $\eta$ mesons, which could be
obtained using the $v_2$ for charged pions and kaons. 

We add that a relation~\cite{rani} between the $p_T$ integrated
 anisotropy for $\pi^0$ and 
photons from its decay, obtained using simulations, 
 has been successfully used to analyze the 
azimuthal anisotropy of decay photons in the WA98 experiment~\cite{wa98_pmd}.

Thus, an experimental verification of the empirical findings
for the transverse momentum dependence of the elliptic flow 
reported in the present work, may be of considerable interest.

 In general,
such a study would also account for  the acceptance and efficiency of the 
photon detectors, which is beyond the scope of this work.

\end{document}